\documentclass[conf]{new-aiaa}
\usepackage[utf8]{inputenc}

\usepackage{graphicx}
\usepackage{amsmath}
\usepackage[version=4]{mhchem}
\usepackage{siunitx}
\usepackage{longtable,tabularx}
\newcommand{\abs}[1]{\lvert #1 \rvert}
\newcommand{\norm}[1]{\lVert #1 \rVert}
\newcommand{\grad}[2]{{\rm grad}_{#2}\left( #1 \right)}

\newcommand{\cartu}[1]{x^{(#1)}}
\newcommand{\cartusq}[1]{x^{(#1)^2}}
\title{Sensitivity computation of statistically stationary 
quantities in turbulent flows}

\author{Nisha Chandramoorthy\footnote{Ph.D. Candidate, Mechanical Engineering and Computation, Bldg. 37, AIAA Student Member.} and Qiqi Wang\footnote{Associate Professor, Aeronautics and Astronautics, Bldg. 33, AIAA Associate Fellow}}
\affil{Massachusetts Institute of Technology, Cambridge, MA, 02139}
\begin{document}

\maketitle

\begin{abstract}
It is well-known that linearized perturbation methods for sensitivity analysis, such as tangent 
or adjoint equation-based, finite difference and automatic differentiation are not suitable 
for turbulent flows. The reason is that turbulent flows exhibit chaotic dynamics, leading to the norm 
of an infinitesimal perturbation to the state growing exponentially in time. As a result, 
these conventional methods cannot be used to compute the derivatives of long-time averaged
quantities to control or design inputs. The ensemble-based approaches \cite{lea,eyink}
and shadowing-based approaches (\cite{qiqi,angxiu,patrick}) to circumvent the problems
of the conventional methods in chaotic systems, also suffer from computational 
impracticality and lack of consistency guarantees, respectively. We 
introduce the space-split sensitivity, or the S3 algorithm, which 
is a Monte Carlo approach to the chaotic sensitivity computation problem. In this 
work, we derive the S3 algorithm under simplifying assumptions on the dynamics and 
present a numerical validation on a low-dimensional example of chaos. 
\end{abstract}

\section{Nomenclature}

{\renewcommand\arraystretch{1.0}
\noindent\begin{longtable*}{@{}l @{\quad\quad} l@{}}
$d$ & phase space dimension \\
$u$ & a phase point or a state vector in $\mathbb{R}^d$ \\
$u_n$ & state vector at time $n$ starting from an initial state $u_0$ \\
$F(u,s)$ & a parameterized discrete-time system representing chaotic dynamics; \\
	&The parameter is $s$\\
$J$ & a scalar objective function of interest \\
$\mu^s$ & an ergodic, invariant probability distribution of SRB type \\
$\langle J, (\partial\mu^s/\partial s)\rangle$ & required sensitivity \\
	$F^i_{uj}$ & the product of Jacobians at $u_{j+i-1},u_{j+i-2},\cdots,u_{j}$, in this order: 
$F_{u(j+i-1)}F_{u(j+i-2)}\cdots F_{uj}$; \\
	&the identity operator when $i=0$ \\ 
$d_u, d_s$ & dimension of the unstable and stable subspaces respectively \\
$E^{\rm u}(u), E^{\rm s}(u)$ & unstable and stable tangent subspaces at a phase point $u$, respectively \\
$V^k_n$ & $k$-th tangent covariant Lyapunov vector (CLV) corresponding to the \\
	&Lyapunov exponent (LE) $\lambda^k$ at the phase point $u_n$ \\
$W^k_n$ & $k$-th adjoint CLV corresponding to the LE $\lambda^k$ at the phase point $u_n$ \\
$N'$ & number of timesteps for time correlation decay \\
$N$ & number of samples or length of primal trajectory \\

\end{longtable*}}
\section{Introduction}
\label{sec:introduction}
With recent advances in large-scale computing, high-fidelity simulations, 
such as large eddy simulations or direct numerical simulations of turbulent flow,
are becoming increasingly common in aerodynamic modelling. Eddy-resolving simulations 
are predicted to become critical in the near future for design of next-generation 
aero-engines and turbomachinery wherein there is a greater demand for reduction in 
noise and pollutant emissions and higher efficiency \cite{dnsAirfoil,tyacke,yang}. 
Such examples call for an in-depth resolution of transition phenomena, shear instabilities, heat transfers,
separations and so on, that require DNS/LES. Sensitivity analysis in these high-fidelity simulations 
is as yet undeveloped. In comparison, gradient information on RANS simulations and non-chaotic Navier-Stokes simulations have 
greatly benefited uncertainty quantification \citep{palacios, qiqi-unstart}, mesh adaptation \citep{fidkowski}, flow control
\cite{rizzetta}, noise reduction \citep{bodony, engblom, paul}  and aerostructural design
optimization applications \citep{peter,giles1,giles2, alonso2, nielsen}. The gradients are 
computed in these simulations in the traditional way -- from the solution of tangent or 
adjoint equations. 

In the context of sensitivity analysis, chaotic simulations pose unique challenges. Since 
the instantaneous quantities in chaotic systems are stochastic in nature, meaningful objectives 
for design optimization and other applications are long-time averages of observables such as 
lift, drag, pressure loss, heat transfer etc. To compute the derivatives of long-time averages to 
changes in system inputs, one would conventionally use tangent or adjoint methods. But, the solution 
of linearized perturbation equations (including those obtained, with no numerical error, through automatic 
differentiation for instance) all have exponentially increasing norms in time -- unbounded growth of infinitesimal perturbations 
to the state is a property of chaotic dynamics. An early approach, known as ensemble sensitivity 
computation, due to Lea {\emph et al.} \cite{lea}, proposed to take the sample average of short time 
sensitivities computed using these linearized perturbation equations. However, the ensemble sensitivity
approach is computationally intractable, as some studies have shown \cite{nisha_ES, eyink}. The more 
recent shadowing-based approaches \cite{angxiu, angxiu_nilsas, angxiu_jfm, qiqi, patrick} tackle this 
problem by numerically computing the shadowing perturbation solution, a carefully constructed tangent 
or an adjoint solution that remains bounded over a long time window. These approaches 
have been successfully applied to 3D flows over a cylinder \cite{angxiu_jfm}. However, there are 
some examples \cite{patrick_KS} where the approach has been shown to not converge to the correct 
sensitivity. In this work, we propose an alternative algorithm -- space-split 
sensitivity (S3) -- to obtain the sensitivity of infinite-time averages or equivalently, statistical 
averages in chaotic systems.

In the next section, we describe some mathematical prerequisites for the S3 algorithm. In section 
\ref{sec:derivation}, we derive the algorithm for the simplistic case of one unstable Lyapunov 
exponent. Section \ref{sec:algorithm} presents the algorithm in full and numerical results 
validating S3 on a low-dimensional example of chaos is discussed in section \ref{sec:kuznetsov}.

\section{Problem Setup}
\label{sec:setup}
The primal problem is the numerical simulation 
of a parameterized turbulent fluid flow. We assume it is of 
the form,
\begin{align}
	u_{n+1} = F(u_n, s), \quad n = 0,1,\cdots, \;\; u_n \in \mathbb{R}^d.
	\label{eqn:primal}
\end{align}
With $u_0$ as an initial flow field, the function $F$ takes in $u_n$, 
the flow field at timestep $n$, to produce the flow field at the next timestep 
$n+1$. The function $F$ represents the spatially and temporally discretized
flow solver, with $s \in \mathbb{R}$ being a scalar parameter of the flow (e.g., inlet 
Mach number, a shape parameter of an airfoil etc). We use $J$ to denote a 
scalar objective function of interest, e.g., lift/drag in a flow over an airfoil,
pressure loss in a turbine wake etc. The observed value at timestep $n$ of $J$
is denoted as $J_n$. The primal simulation resolves chaotic timescales. 
Consequently, a time-average of $J$, given by $\langle J\rangle_N := (1/N)\sum_{n=0}^{N-1} J_n$
that is observable and is of engineering interest (in e.g., design and 
optimization) corresponds to a large $N$. We are interested in the sensitivity of 
the infinite-time average of $J$ to a small change in $s$. That is, 
we want to compute the quantity $d(\lim_{N\to \infty}\langle J\rangle_N)/ds$. 
\section{Mathematical background}
\label{sec:background}
The goal of this section is to present some mathematical prerequisites
for the derivation of the S3 algorithm. The present treatment is not meant 
to be rigorous (the reader is referred to \cite{katok} for a detailed 
exposition) but rather a brief, intuitive explanation of some concepts 
from dynamical systems and ergodic theory, to the extent applicable to 
the developments in the further sections.
\subsection{Phase space perspective}
The flow field at time $n$, $u_n$, is a point in a 
$d$-dimensional phase space, denoted by ${\cal M}$. 
Each dimension corresponds to a degree of freedom in the flow simulation. 
For instance, in a 3D incompressible flow where we solve for the 3 velocity components and the pressure at 
every grid point, the dimension of the phase space $d$ is equal to 
$4$ times the number of grid points. Each unsteady flow simulation 
sequentially visits a finite number of points in ${\cal M}$ -- a finite 
length trajectory in phase space. The set of all trajectories in phase 
space converges to a closed and bounded set in $\mathbb{R}^d$. This set,
called an attractor, contains the asymptotic trajectories of all points 
starting from initial conditions in the neighbourhood of the set. 

We are interested in statistically stationary flows i.e., fluid flows 
in which the state vector has achieved a time-invariant probability 
distribution on the attractor. That is, each trajectory (flow simulation) 
samples a finite number of phase points (state vectors) distributed 
according to an invariant probability distribution. An ensemble average 
refers to taking an expected value according to this distribution; thus,
it is the operation of taking the sample average at phase points on the attractor 
in the limit as the number of samples goes to infinity. 

\subsection{Linearized perturbation equations and uniform hyperbolicity}
\label{sec:lyapunov}
The tangent space at a phase point $u$, denoted by $T_u {\cal M}$, is a $d$-dimensional 
vector space, that consists of all possible infinitesimal perturbations to the state. Each tangent 
vector gives a direction of perturbation; given a scalar function on the phase space, 
the tangent vector provides a valid direction to compute the directional derivative of the scalar function along.
In this paper, we deal with a class of idealized dynamical systems,
called uniformly hyperbolic systems. In uniformly hyperbolic
systems, the tangent space at every point can be decomposed into stable 
and unstable, lower dimensional subspaces, denoted by $E^{\rm s}$ and $E^{\rm u}$
respectively -- $T_u {\cal M} = E^{\rm u}(u) \oplus E^{\rm s}(u)$. It is important 
to note that this decomposition is a direct sum decomposition, not an
orthogonal decomposition. Roughly speaking, uniform hyperbolicity is the existence 
of a stable-unstable decomposition of the tangent space on the attractor that 
is covariant (i.e., stable (unstable) vectors are mapped to stable (unstable) 
vectors under the tangent dynamics along a primal trajectory) with the dynamics. 

We will now introduce the tangent dynamics, which essentially track the evolution 
of infinitesimal perturbations linearized about a primal trajectory. The familiar tangent equation 
for the evolution of an infinitesimal perturbation to the parameter, $\zeta_n := 
\frac{\partial u_n}{\partial s}$ (at a fixed primal initial condition $u_0$), is 
given by,
\begin{align}
	\label{eqn:tangent}
	\zeta_{n+1} &= F_{un} \zeta_n + F_{sn}, \\
	\zeta_0 &= 0 \in \mathbb{R}^d,
\end{align}
where $F_{un}$ is the Jacobian matrix consisting of the partial derivatives 
of the form $\partial F^{(i)}/\partial u^{(j)}$, evaluated at the phase point 
$u_n$; the superscript $(i)$ indicates the i-th Cartesian coordinate. Following 
a similar notation, $F_{sn}$ indicates the partial derivative $\partial F/\partial s$,
evaluated at the phase point $u_n$, at a reference value of the parameter $s$.
Note that $F_{sn}$ is a tangent vector belonging to the tangent space at $u_{n+1}$, 
$T_{u_{n+1}}{\cal M}$; this is clear from Eq.\ref{eqn:tangent} where it can be subtracted 
from $\zeta_{n+1}$, a tangent vector in $T_{u_{n+1}} {\cal M}$. We denote by $X_n$ 
the tangent vector of the source term in the previous timestep, i.e., $X_n := F_{s(n-1)}
\in T_{u_n}{\cal M}$. 

The homogeneous tangent equation gives the evolution of an infinitesimal perturbation 
to the initial state. That is, let $\zeta_0 \in T_{u_0} {\cal M}$
and $u_0' = u_0 + \epsilon \zeta_0$. Then, keeping $s$ fixed at a reference value,
\begin{align*}
	\zeta_n := \partial u_n/\partial \epsilon = \lim_{\epsilon \to 0} 
	\frac{u_n' - u_n}{\epsilon},
\end{align*}
where $u_n'$ is the state at time $n$ starting with $u_0'$ as the initial condition.
The tangent vector $\zeta_n$ satisfies the following tangent equation,
\begin{align}
	\label{eqn:homogeneousTangent}
	\zeta_{n+1} = F_{un} \zeta_n,  
\end{align}
which is called the homogeneous tangent equation since it is the conventional 
tangent equation with the source term zero. We are now ready to define uniform 
hyperbolicity more precisely using the behavior of linearized perturbation solutions. 
To wit, in uniformly hyperbolic systems, there exist constants $C > 0$ and 
$\lambda \in (0,1)$ such that for every initial condition $u_0$, if $\zeta_0 
\in E^{\rm s}(u_0)$, then, the solution of the homogeneous tangent equation 
(Eq.\ref{eqn:homogeneousTangent}), $\zeta_n \in E^{\rm s}(u_n)$ and $\norm{\zeta_n} \leq 
C \lambda^n \norm{\zeta_0}$, for all $n > 0$. Similarly, if $\zeta_0 
\in E^{\rm u}(u_0)$, then, the solution of the homogeneous tangent equation 
(Eq.\ref{eqn:homogeneousTangent}) backward in time (using the inverse 
Jacobian $F_{un}^{-1}$), $\zeta_{-n} \in E^{\rm u}(u_{-n})$ and $\norm{\zeta_{-n}} \leq 
C \lambda^n \norm{\zeta_0}$, for all $n >0$.

Suppose $J$ is a scalar-valued state 
function. The conventional adjoint equation gives the backward evolution of the
derivative of $J$ at time $N$, denoted as $J_N$, to an infinitesimal perturbation 
to the state $u_n$. The familiar adjoint equation for $\xi_n := \partial J_N/\partial u_n$,
is given by,
\begin{align}
	\label{eqn:adjoint}
	\xi_n &= F_{un}^T \xi_{n+1} + \grad{J_n}{n},\;\; n=N,N-1,\cdots, \\
	\xi_{N+1} &= 0 \in \mathbb{R}^d,
\end{align}
where $\grad{J_n}{n}$ is the gradient of the function $J$ evaluated at 
$u_n$. Setting the source term of the above equation to zero, we get the 
following homogeneous adjoint equation,
\begin{align}
	\label{eqn:homogeneousAdjoint}
	\xi_n &= F_{un}^T \xi_{n+1}.   
\end{align}
In a uniformly hyperbolic system, we can also write down a different 
decomposition of the tangent space at a phase point classifying tangent 
vectors that decay or grow exponentially in norm, under the homogeneous 
adjoint dynamics (Eq.\ref{eqn:homogeneousAdjoint}). This decomposition 
is related to the decomposition into $E^{\rm u}$ and $E^{\rm s}$. If 
$\xi_n$ is a stable adjoint vector, that is, 
$\xi_n \in T^*_{u_n} {\cal M}$ is such that $\norm{\xi_0} \leq C 
\lambda^n\norm{\xi_n}$, then, $\xi_n \in E^{{\rm u}^{\perp}}(u_n)$.  
Similarly, one can show that unstable adjoint vectors are also covariant
and belong to the orthocomplement of the stable tangent subspaces,
$E^{{\rm s}^\perp}$. Thus, the decomposition based on stable and 
unstable adjoint vectors at a phase point is given by $T^*_u {\cal M} 
= E^{{\rm u}^\perp}(u) \oplus E^{{\rm s}^\perp}(u)$. 
\subsection{Covariant Lyapunov vectors}
Covariant Lyapunov vectors (CLVs), denoted by $V^1,\cdots,V^d$ form a special non-orthogonal  
basis for the tangent space at every point. Here $V^1,\cdots,V^d$ denote the normalized basis
vectors. The CLV basis is covariant in the sense that 
if $\zeta_0 := V^k_0 \in T_{u_0}{\cal M}$ is an initial condition of the homogeneous tangent equation 
(Eq.\ref{eqn:homogeneousTangent}), then, the solution at time $n$, $\zeta_n \in {\rm span}\{V^k_n\}$, for 
each $k=1,\cdots, d$. Moreover, the asymptotic logarithmic growth rates of the CLVs are independent of the primal 
initial condition and are the so-called Lyapunov exponents. Equivalently, Lyapunov exponents 
can also be expressed as the ensemble average of the logarithmic growth rate of the CLVs. That is,
the $k$-th Lyapunov exponent (LE), denoted by $\lambda^k$ is the ensemble average over initial conditions 
$u_0$ of, $\log(\norm{F_{u0} V^k_0}/\norm{V^k_0}) = \log(\norm{F_{u0} V^k_0})$. We assume that 
no two LEs are equal and index them in the descending order as, $\lambda^1 > \lambda^2 > 
\cdots > \lambda^k$. The adjoint CLVs have an analogous definition with the same LEs and the 
homogeneous adjoint equation replacing the role of the 
homogeneous tangent equation. We denote the normalized adjoint CLVs using $W^1,\cdots, W^d$. 

Furthermore, the tangent CLVs at $u$ corresponding to unstable homogeneous tangent solutions,
i.e., $\lambda^k > 1$, form a basis for $E^{\rm u}(u)$. If $d_u < d$ is the dimension of 
the unstable subspace (which is the same at every phase point) and $d_s = d - d_u$ is the 
dimension of the stable subspace, then, $E^{\rm u}(u) = {\rm span}\{
V^1(u), V^2(u), \cdots, V^{d_u}(u)\}$ and similarly, $E^{\rm s}(u) = {\rm span}\{
V^{d_u+1}(u), \cdots, V^{d}(u)\}$. The orthocomplements, $E^{{\rm u}^\perp}$ and $E^{{\rm s}^\perp}$,
are spanned by the adjoint stable CLVs ($W^k, k = d_u + 1, \cdots, d$) and adjoint unstable CLVs
($W^k, k = 1, \cdots, d_u$) respectively. 
\subsection{Tangent-adjoint orthogonality property}
\label{sec:tao}
Since the CLVs are tangent vectors, the orthogonality property of stable and unstable 
adjoint and tangent vectors, alluded to in section \ref{sec:lyapunov}, applies to them. 
More explicitly, $W^k \in E^{{\rm s}^\perp}$ for $k=1,\cdots, d_u$ and $V^j \in E^{\rm s}$
for $j=d_u +1, \cdots, d$ and so, $W^k \cdot V^j = 0$, at every phase point. In words, 
a stable tangent CLV is orthogonal to an unstable adjoint CLV at every point. Similarly,
an unstable tangent CLV is orthogonal to a stable adjoint CLV. This orthogonality will 
be referred to as the tangent-adjoint orthogonality or TAO property. 

\subsection{Measure preservation}
In statistically stationary systems, the assumption of uniform hyperbolicity guarantees 
the existence of a unique invariant probability distribution called the SRB distribution 
(see \cite{young} for an introduction to SRB measures), that has certain convenient properties. 
Most importantly, it is a physically observable distribution, 
in the sense that, long time averages of functions along trajectories converge to their expected value according to the 
SRB distribution. That is, $\lim_{N\to \infty}\langle J\rangle_N := \int J \; d\mu^s$, where $\mu^s$ is 
the SRB distribution, for almost every initial condition $u_0$. The superscript $s$ indicates the SRB distribution depends on the 
parameter $s$. The sensitivity we would like to compute can therefore be equivalently 
written as $d/ds (\lim_{N\to \infty} \langle J\rangle_N) = d/ds (\int J \; d\mu^s) = 
\int J \; d\mu^s/ds$.

Denoting the integral pairing $\int J d\mu^s$ using $\langle J, \mu^s\rangle$, $\langle J_0, \mu^s\rangle = 
\langle J_n, \mu^s\rangle$, for any $n$, where, with a slight abuse of notation, $J_i$ denotes the 
function composition $J \circ F \circ (i\; {\rm times})\;\circ F$. That is, considering 
an ensemble of initial conditions, independent and identically distributed according to $\mu^s$, 
the average of a function $J$ over the primal trajectories 
starting from the ensemble, is independent 
of time. This statement essentially captures the statistical stationarity of the fluid 
flow with respect to the SRB distribution. In another interpretation, we can arbitrarily 
move the time origin to any time while computing ensemble averages with respect to an 
invariant distribution such as the SRB distribution. This property is called measure preservation 
property or MPP for short.

\section{Derivation of the S3 algorithm} 
\label{sec:derivation}
In this section, we derive the S3 algorithm for a uniformly hyperbolic 
chaotic system, placing
emphasis on the physical/computational motivation for each step;
the rigorous mathematical details of some steps of the derivation 
are left to the appendix. From Ruelle's linear response theory 
(see \cite{ruelle,ruelle1} for the derivation) the 
sensitivity of interest can be expressed through the following 
formula, where $X(u_n) = X_n := F_{s(n-1)}$, 
\begin{align}
	\notag
	\langle J, \dfrac{\partial \mu^s}{\partial s}\rangle 
	&= \sum_{n'=0}^\infty 
	\langle \grad{J_{n'}}{0} \cdot X, \mu^s\rangle.
\end{align}
We decompose the formula by writing the tangent vector 
$X = X^{\rm s} + X^{\rm u}$, as a sum of its stable 
and unstable components, i.e., $X^{\rm s} \in E^{\rm s}$ 
and $X^{\rm u} \in E^{\rm u}$ at every phase point. Then,
\begin{align}
	\label{eqn:splitRuelleFormula}
	\langle J, \dfrac{\partial \mu^s}{\partial s}\rangle 
	&= \sum_{n'=0}^\infty \langle \grad{J_{n'}}{0} \cdot X^{\rm s}, \mu^s\rangle + 
	\sum_{n'=0}^\infty  \langle \grad{J_{n'}}{0} \cdot X^{\rm u}, \mu^s\rangle. 
\end{align}
We refer to the two components of Eq.\ref{eqn:splitRuelleFormula} as the 
stable and unstable contributions, and denote them using the subscripts 
``stable'' and ``unstable'' respectively. The stable contribution can be 
computed using the conventional tangent solution approach, 
as in a non-chaotic system. To see this, note that in the stable contribution,
the summation and the ensemble-averaging operations can be commuted, since 
at almost every initial condition, the instantaneous sensitivity is bounded at all 
times. That is, the stable contribution can be written as 
\begin{align}
	\langle J, \frac{\partial \mu^s}{\partial s} \rangle_{\rm stable} 
	&=  \sum_{n'=0}^\infty  \langle \grad{J_{n'}}{0}\cdot
	X^{\rm s},\mu^s \rangle = 
	\langle \sum_{n'=0}^\infty \grad{J_{0}}{-n'}\cdot
	X^{\rm s}_{-n'},\mu^s \rangle,
\end{align}
where the second equality is obtained by applying the MPP. Further, using 
the fact that the homogeneous tangent solution is stable for all time in this 
case, 
\begin{align}
	\label{eqn:stableContributionMPP}
	\langle J, \frac{\partial \mu^s}{\partial s} \rangle_{\rm stable} 
	&= \sum_{n'=0}^\infty 
	\langle (F_{u(-n')}^{n'})^T \grad{J_0}{0} \cdot X^{\rm s}_{-n'}, \mu^s\rangle = 
	\sum_{n'=0}^\infty 
	\langle \grad{J_0}{0} \cdot  F_{u(-n')}^{n'} X^{\rm s}_{-n'}, 
	\mu^s \rangle \\
	\label{eqn:stableContributionSummationInside}
	&= \langle \grad{J_0}{0} \cdot \sum_{n'=0}^\infty 
	F_{u(-n')}^{n'} X^{\rm s}_{-n'}, \mu^s \rangle  =
	\langle \grad{J_0}{0} \cdot \psi_0^{\rm s}, \mu^s \rangle,
\end{align}
where $\psi_0^{\rm s}$ is the series summation of the homogeneous tangent 
solutions. In practice, it can be computed from solving a stable inhomogeneous 
tangent equation, the conventional tangent equation with a stable source term at every timestep. 
Explicitly, suppose $\zeta_n^{\rm s}$ is the solution to the following equation,
\begin{align}
	\notag
	\zeta_{n+1}^{\rm s} &= F_{un} \zeta_{n}^{\rm s} + X^{\rm s}_{n+1},\;\;n=0,1,\cdots,\\
	\label{eqn:stableTangentEquation}
	\zeta_0^{\rm s} &= X_0^{\rm s} \in \mathbb{R}^d.
\end{align} Then, the stable contribution can be written approximately as the following 
ergodic average,
\begin{align}
\label{eqn:stableContributionTangentApproximation}
	\langle J, \frac{\partial \mu^s}{\partial s} \rangle_{\rm stable} 
	&\approx \frac{1}{N}\sum_{n=0}^{N-1}
	\grad{J_n}{n} \cdot \psi_n^{\rm s} \approx  \frac{1}{N}\sum_{n=0}^{N-1}
	\grad{J_n}{n} \cdot \zeta_n^{\rm s}. 
\end{align} The approximation $\psi_n^{\rm s} \approx \zeta_n^{\rm s}$ 
in Eq.\ref{eqn:stableContributionTangentApproximation} gets better with $n$. 
From an explicit expression for the stable tangent solution $\zeta_n^{\rm s} = \sum_{n'=0}^{n}
F_{u(n-n')}^{n'} X^{\rm s}_{(n-n')}$, we can see that it does not include the terms of the sequence 
$\left\{ F_{u(n-n')}^{n'} X^{\rm s}_{(n-n')}\right\}, n'=0,1,\cdots$ for $n'>n$. But, since 
the sequence is exponentially decreasing in norm, the truncation error is insignificant at
large $n$. Thus, the computation of 
the stable contribution resembles sensitivity computation in a non-chaotic system.

In deriving the computation of the unstable contribution, we now make the simplifying assumption 
that the unstable subspace is one-dimensional, i.e., $d_u=1$. Then the unstable contribution can 
be written as follows,
\begin{align}
	\label{eqn:unstableContribution1DU}
	\langle J, \dfrac{\partial \mu^s}{\partial s}\rangle_{\rm unstable}
	= \sum_{n'=0}^{\infty} \bigg(
	\langle x_0 \grad{J_{n'}}{0}\cdot V_0^1, \mu^s \rangle \bigg),  
\end{align}
where the scalar field $x$ is the component of $X^{\rm u}$ along $V_0^1$. That is,
$X^{\rm u}_n = x_n V_n^1$, along a trajectory. We seek an indirect computation of each 
summand that does not involve taking the derivative of $J_{n'}$. 
We derive such a computation by establishing an iterative formula 
that is stable in the sense that its rate of convergence is uniformly 
bounded over $n'$. The reader is referred to the Appendix section 
\ref{sec:unstableContributionFormula} for all the steps of the derivation.
Here we will describe the mathematical intuition for why such a formula exists 
and how to compute it.

Roughly speaking, we argue that the computational constraints we have 
imposed direct us to an efficient approach. In particular, 
we seek an algorithm that scales linearly with $N$, the number 
of samples or the length of a trajectory used for a Monte Carlo
approximation. Now note that each $n'$ summand in the unstable contribution is 
a linear functional (an operation on a function that returns a scalar) on a suitable Hilbert space 
containing $J_{n'}$. As a result of Riesz representation theorem, each $n'$ summand 
has a representation as an inner product (on a Hilbert space 
of functions) of the objective function with another state function.  
That is, $\langle \grad{J_{n'}}{0} \cdot X_0^{\rm u}, \mu^s \rangle 
= \langle J_{n'} G^{\rm u}_0, \mu^s\rangle$, where $G^{\rm u}$ is a 
scalar function and the right hand side of the equality is by definition the inner 
product. We will henceforth refer to $G^{\rm u}$ 
as the Riesz representation of unstable ensemble derivative (or RUED, for short).
Note that the RUED does not capture the pointwise unstable derivative but rather 
its ensemble average, which is bounded for each $n'$. The alternative representation
afforded by the RUED is preferable for computation because it is able to show a Monte Carlo 
convergence as an ergodic average, as stipulated. This claim can be reasoned as follows. 
Suppose we have computed the RUED, $G^{\rm u}$, that satisfies $\langle J_{n'} G^{\rm u}_0, 
\mu^s \rangle = \langle \grad{J_{n'}}{0} \cdot X_0^{\rm u}, \mu^s\rangle$. Then, the 
unstable contribution, upon applying the MPP, can be written as,
\begin{align}
	\label{eqn:timeCorrelation}
	\langle J, \frac{d\mu^s}{d s}\rangle_{\rm unstable} = 
	\sum_{n'=0}^\infty \langle J_{0} G^{\rm u}_{-n'}, \mu^s \rangle. 
\end{align}
Note that the $n'$ summand is the time correlation at $n'$ between the two 
state functions, $J$ and $G^{\rm u}$. State functions in uniformly hyperbolic 
systems, in general, enjoy an exponential decay of time correlations, i.e., 
Eq.\ref{eqn:timeCorrelation} converges exponentially with $n'$ to the  
product of the ensemble averages of $J$ and $G^{\rm u}$. Therefore,
in practice, when we compute the above expression (Eq.\ref{eqn:timeCorrelation})
as an ergodic average, we can make an additional approximation,
\begin{align}
	\label{eqn:timeCorrelationErgodicAverage}
	\langle J, \frac{d\mu^s}{d s}\rangle_{\rm unstable} \approx 
	\frac{1}{N} \sum_{n'=0}^\infty \sum_{n=0}^{N-1} 
	J_{n} G^{\rm u}_{n-n'} \approx  
	\frac{1}{N} \sum_{n=0}^{N-1} 
	J_{n} \sum_{n'=0}^{N'-1}G^{\rm u}_{n-n'}, 
\end{align}
where $N'$ represents a timescale for the decay of correlations. Since 
$N'$ is expected to be small in comparison to $N$, this results in 
a computation that scales roughly linearly with $N$, as we stipulated. Moreover,
we expect the central limit theorem to be valid for the above $N$-sample estimator.
This is in turn due to the fact that the variance in the random variable 
$J_{n'} G^{\rm u}_0$ (the randomness comes from the initial condition) 
is uniformly bounded over $n'$. In contrast, the variance of the integrand in 
the original form (Eq.\ref{eqn:unstableContribution1DU}), 
$\grad{J_{n'}}{0} \cdot X_0^{\rm u}$, exhibits unbounded growth with $n'$. 
Thus, an $N$-term ergodic average of Eq.\ref{eqn:unstableContribution1DU} shows 
a (much) slower convergence than the $1/\sqrt{N}$ convergence predicted by the central 
limit theorem (see \cite{nisha_ES} for an analysis of ensemble estimates 
of the gradient term). 

The question remains as to how to compute the RUED, and is partially 
answered by integration by parts. Motivated to avoid the computation of a term 
such as $\grad{J_{n'}}{0}$, which is exponentially increasing in norm with $n'$, 
we rewrite the unstable contribution as follows,
\begin{align}
	\label{eqn:unstableContribution1DU}
	\langle J, \dfrac{\partial \mu^s}{\partial s}\rangle_{\rm unstable}
	=  \sum_{n'=0}^{\infty} \bigg(
	\langle \grad{x_0 J_{n'}}{0}\cdot V_0^1, \mu^s \rangle -   
	\langle J_{n'} \grad{x_0}{0}\cdot V_0^1, \mu^s \rangle \bigg).  
\end{align}
The above statement amounts to integration by parts which, as we expect, 
has a regularization effect. We can see that the second term above, which 
is in the form of a time correlation between the state functions $J$ and 
$\grad{x}{\cdot}\cdot V_{\cdot}^1$, leads to a Monte Carlo method. 
We are now left with expressing the first term also as a time correlation 
in order to determine $G^{\rm u}$ completely. In the Appendix section 
\ref{sec:unstableContributionFormula}, we determine 
a scalar field $\alpha^{\rm u}$ that satisfies $\langle \grad{x_0 J_{n'}}{0}
\cdot V_0^1, \mu^s \rangle = \langle J_{n'} \alpha^{\rm u}, \mu^s \rangle$.   
By using the MPP and algebraic manipulations, we arrive at 
the following formula for $\alpha^{\rm u}$ (see Appendix section 
\ref{sec:unstableContributionFormula} for the full derivation),
\begin{align}
\label{eqn:ruedExpression}
	\alpha^{\rm u}_n &= x_n \sum_{n'=-\infty}^{n-1}
	\frac{\phi_{n'}}{\Pi_{i=n'+1}^{n-1}\norm{F_{ui}V_i^1}}, 
\end{align}
where $\phi_n := -\grad{1/\norm{F_{un} V_n^1}}{n}$. One can see that 
$\alpha^{\rm u}_n \approx \beta^{\rm u}_n$, the solution of the following equation,
\begin{align}
	\label{eqn:ruedEquation}
	\beta^{\rm u}_{n+1} &= \frac{\beta^{\rm u}_n x_{n+1}}{\norm{F_{un}V_n^1} x_n} + 
	x_{n+1} \phi_{n},\;\;
	n=0,1,2,\cdots\\ 
	\beta^{\rm u}_0 &= \phi_{-1} x_0.
\end{align}
The above equation will be referred to as the RUED equation, since it 
computes the RUED term partially. Together with the contribution from the 
second term in Eq.\ref{eqn:unstableContribution1DU}, we have the following 
computable formula for the RUED term, 
\begin{align}
	G^{\rm u}_n = \alpha^{\rm u}_n - \grad{x_n}{n} \cdot V_n^1.
    \label{eqn:ruedTermComplete}
\end{align}
The unstable contribution can then be expressed as the following ergodic average 
approximation of the time correlation between $J$ and $G^{\rm u}$,
\begin{align}
	\label{eqn:overallUnstableContribution}
	\langle J, \frac{\partial \mu^s}{\partial s}\rangle_{\rm unstable} 
	\approx \frac{1}{N} \sum_{n=0}^{N-1} J_n \sum_{n'=n-N'+1}^n G^{\rm u}_{n'}.
\end{align}
Finally, combining the stable and unstable contributions, we obtain the 
overall sensitivity,
\begin{align}
	\label{eqn:SensitivityEfficient}
	\langle J, \dfrac{\partial \mu^s}{\partial s}\rangle_{\rm unstable} + 
	\langle J, \dfrac{\partial \mu^s}{\partial s}\rangle_{\rm stable}  
	&\approx \frac{1}{N} \sum_{n=0}^{N-1}
	\Bigg(  J_n \sum_{n'=n-N'+1}^n G^{\rm u}_{n'}
	+ \grad{J_n}{n} \cdot \zeta_n^{\rm s} \Bigg). 
\end{align}
\section{The S3 algorithm}
\label{sec:algorithm}
In this section, we describe the implementation of 
the computable formula in Eq.\ref{eqn:SensitivityEfficient}
that was derived in the previous section \ref{sec:derivation}.
We detail the following stepwise procedure that results in the 
sensitivity of interest $\langle J, ({\partial \mu^s}/{\partial s})\rangle$. 
\begin{enumerate}
		\item Primal and unstable CLV computation
		\begin{enumerate}
				\item[1.1]  Choose a random initial condition $u_0$ and solve the primal problem 
				in Eq.\ref{eqn:primal} for a large ${\cal N}$, producing the trajectory $u_0, u_1, 
				\cdots, u_{\cal N}$. 
		\item[1.2] Solve the homogeneous tangent equation \ref{eqn:homogeneousTangent} along 
				the computed primal trajectory, by starting with a random initial condition $\zeta_0$, 
				and normalizing the solution at each timestep. The resulting unit vectors  
				$\zeta_n$ approximate the unstable CLVs $V^1_n$ accurately after some 
				time $N_{\rm tan}$. 
		\item[1.3] Solve the homogeneous adjoint equation \ref{eqn:homogeneousAdjoint}
				backward in time, again dividing the adjoint solution by its norm at each timestep, 
				starting with a random initial condition $\xi_{\cal N}$. 
				The resulting unit solution vectors $\xi_n$ accurately approximate 
				the unstable adjoint CLVs $W^1_n$ after performing this procedure for time 
				$N_{\rm adj}$, i.e., $W^1_n$ are accurate for all $n \leq {\cal N}-N_{\rm adj}$. 
		\item[1.4] Shift the time origin to $N_{\rm tan}$. We have obtained the unstable tangent and 
				adjoint CLVs $V_n^1, W_n^1$ respectively, along the trajectory $u_n,\;
				n = 0,1,\cdots,{\cal N}-N_{\rm tan}
				- N_{\rm adj}$.
		\end{enumerate}
		\item Stable-unstable pertubation splitting and the RUED term pre-computations 
		\begin{enumerate}
				\item[2.1] Compute $c^1_n := W^1_n\cdot V^1_n$, for all $n=0,1,\cdots,{\cal N}- N_{\rm tan} 
				- N_{\rm adj}$. 
		\item[2.2] Compute $X_{n+1} = F_{sn}$ at each $n$ analytically if possible, or through finite difference. 
				Also obtain $x_n = X_n \cdot W_n^1/c_n^1$. Then, obtain the stable-unstable decomposition of $X_n$:
				$X_n^{\rm u} = x_n V_n^1$ and $X_n^{\rm s} = X_n - X_n^{\rm u}$. 
		\item[2.3] Obtain part of the RUED term $-\grad{x_n}{n}\cdot V_n^1$ by using a finite difference 
				approximation of $\grad{x_n}{n}$.
		\item[2.4] Compute $\phi_n := -\grad{1/\norm{F_{un}V_n^1}}{n}$ also as a finite difference approximation. We now 
				have the source term for the RUED equation, $x_{n+1} \phi_n$, at all $n$.
		\end{enumerate}
		\item Initializations for the Monte Carlo loop 
		\begin{enumerate}
				\item[3.1] Choose the times $N_{\rm rued}$ and $N_{\rm tan,s}$ after which the approximation errors
						$\abs{\alpha_n^{\rm u} - \beta_n^{\rm u}}$ and $\norm{\psi_n^{\rm s}-\zeta_n^{\rm s}}$
						corresponding to the RUED (Eq.\ref{eqn:ruedEquation}) and the stable tangent 
						(Eq.\ref{eqn:stableTangentEquation}) equations respectively, are both within specified tolerances. 
						Let the guess $N_{\rm rued}$ be the maximum of the 
						two times. It can be refined as the two equations are solved in the loop to follow.
				\item[3.2] Choose the time $N'$ for the decay of correlations $\langle J G_{-n}^{\rm u}, \mu^s\rangle$. Again, 
						the time $N'$ can be fine-tuned in the following $n$-loop, where the RUED terms $G_n^{\rm u}$ will be  
						obtained. After sufficient iterations, plot on a semilog scale, the time correlation between $J$ and 
						$G^{\rm u}$. If \verb+tol+ is a specified precision and $-\gamma$ and $\log{C}$ are the slope and 
						intercept of the plot respectively, then $N' = (1/\gamma) \log(C/\verb+tol+)$.  
				\item[3.3] The number of samples, $N$, that is to be used for the Monte Carlo estimate 
						of the sensitivity reduces to $N = {\cal N} - N_{\rm tan} - N_{\rm adj} - N_{\rm rued} 
						- N'$.
				\item[3.4] Initialize the loop variable $n$ to $-N_{\rm rued}-N'$. Set zero initial conditions for the stable tangent equation 
						and the RUED equation: $\zeta^{\rm s}_n = 0 \in \mathbb{R}^d$ and 
				$\beta^{\rm u}_n = 0 \in \mathbb{R}$.
		\item[3.5] Initialize an $N'$-long array $q$ to 0. This data structure 
				will be used to compute time correlations for the unstable 
				contribution.
		\item[3.6] Initialize to 0 the stable and unstable 
				sensitivity outputs: $\langle J, (\partial \mu^s/\partial s)\rangle_{\rm stable} 
				=  \langle J, (\partial \mu^s/\partial s)\rangle_{\rm unstable} = 0$. They are to be 
				updated in the following $n$-loop.
		\end{enumerate}
		\item Monte Carlo approximation of the stable and unstable contributions
				\begin{enumerate}
						\item[4.1] Solve the RUED equation (Eq.\ref{eqn:ruedEquation}) for
						the spin-up time $N_{\rm rued}$, to obtain $\beta_{-N'}^{\rm u}$. 
				\item[4.2] Solve the stable tangent equation (Eq.\ref{eqn:stableTangentEquation}) for
						the time $N_{\rm rued} + N'$, to obtain $\zeta_{0}^{\rm s}$. 
				\item[4.3] Continue solving the RUED equation for $n = -N',\cdots,-1$ and set 
						$\alpha_n^{\rm u} = \beta_n^{\rm u}$. Compute $G^{\rm u}_n$ from Eq.\ref{eqn:ruedTermComplete} 
						and step 2.3, for each $n$, and use them to populate $q$. 
						Perform the next steps, 4.4--4.6, for each $n =0,1,\cdots,N-1$. 
				\item[4.4]   Advance the RUED equation by one timestep, obtaining $\beta_n^{\rm u}$. 
				Then set $\alpha_n^{\rm u} = \beta_n^{\rm u}$. Compute $G_n^{\rm u}$ from 
				Eq.\ref{eqn:ruedTermComplete} and using step 2.3, and use it to replace the ($n$ modulo $N'$)-th element of 
				$q$. 
		\item[4.5] Advance the stable tangent equation (Eq.\ref{eqn:stableTangentEquation})
				by one timestep to obtain $\zeta_n^{\rm s}$. Add $\grad{J_n}{n} \cdot \zeta_n^{\rm s}/N$ 
				to the stable contribution, $\langle J, (\partial \mu^s/\partial s)\rangle_{\rm stable}$. 
		\item[4.6] Add the elements of the array $q$ to obtain $q_{{\rm sum},n} = 
				\sum_{n'={n-N'+1}}^n G^{\rm u}_{n'}$. Update the unstable contribution, 
				$\langle J, (\partial \mu^s/\partial s)\rangle_{\rm unstable}$	by adding 
				$q_{{\rm sum},n} J_n/N$.
		\end{enumerate}
		\item[Output:] At the end of the $n$-loop, the overall sensitivity is given by 
				$\langle J, (\partial \mu^s/\partial s)\rangle_{\rm stable} + 
				\langle J, (\partial \mu^s/\partial s)\rangle_{\rm unstable}$.

\end{enumerate}

\section{Example computation}
\label{sec:kuznetsov}
\begin{figure}
	\includegraphics[width=\textwidth]{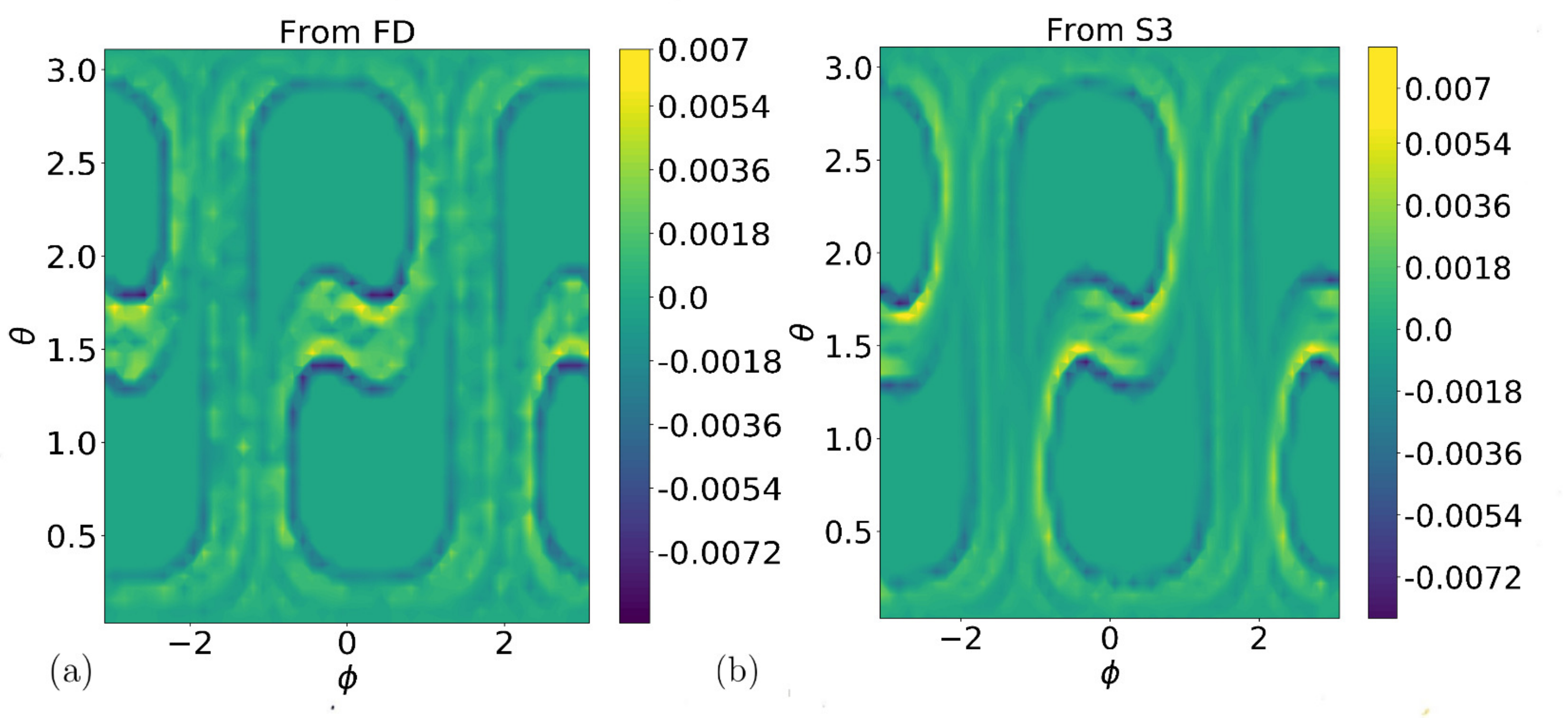}
	\caption{The sensitivity of the ensemble average of nodal basis functions 
	to $s$ in the Kuznetsov-Plykin system. The results obtained from 
	finite difference approximation and from S3 are shown on the left (a) 
	and right (b) respectively.}
\end{figure}
In this section, we consider an example of a three-dimensional uniformly hyperbolic 
chaotic system. The reader is referred to Kuznetsov \cite{kuznetsov}, wherein the 
dynamics is derived as that of a system of non-autonomous coupled oscillators and 
uniform hyperbolicity is numerically verified. Here we are interested in numerical 
results on this system, which we will refer to as the Kuznetsov-Plykin system, of 
the S3 algorithm derived in section \ref{sec:derivation}. We consider the Kuznetsov-Plykin map 
on $u := [x^{(1)},x^{(2)},x^{(3)}]^T \in \mathbb{R}^3$ written in Cartesian coordinates, given by,
\begin{align}
	u_{n+1} = F(u_n,s) = {\mathbf f}_{1,1} \circ \mathbf{f}_{-1,-1}(u_n,s), 
\end{align}
where the function ${\mathbf f_{\cdot,\cdot}}$ is specified in \cite{kuznetsov}. For 
completion, we repeat here the definition of ${\mathbf f}_{\cdot,\cdot}$, where 
the superscript $(k)$ indicates the Cartesian coordinate $k$, 
\begin{align*}
	{\mathbf f}_{\sigma_1,\sigma_2}^{(1)}(u,s) =\: &\sigma_2\: \cartu{3} \\ 
	{\mathbf f}_{\sigma_1,\sigma_2}^{(2)}(u,s) =\: 
	&\frac{ \sqrt{\cartusq{1} + \cartusq{2} }  }
	{\sqrt{\cartusq{1} \:+\: \cartusq{2} 
	\exp{\left(2s( \cartusq{1} \:+\: \cartusq{2})\right)}}} 
	\Big[ \sigma_1 \: \cartu{1} \: 
	 \:\sin{\left(\frac{\pi}{2}
	(\sqrt{2} \cartu{3} + 1 )\right)}  \\
	&+\; \cartu{2} \exp{\left(s(\cartusq{1} + 
	\cartusq{2})\right)}
	\:\cos{\left(\frac{\pi}{2}
	(\sqrt{2} \cartu{3} + 1 )\right)} \Big]  \\
	{\mathbf f}_{\sigma_1,\sigma_2}^{(3)}(u,s) = 
	&\frac{ \sqrt{\cartusq{1} + \cartusq{2} }  }
	{\sqrt{\cartusq{1} \:+\: \cartusq{2} 
	\exp{\left(2s( \cartusq{1} \:+\: \cartusq{2})\right)}}} 
	\Big[ -\sigma_2 \: \cartu{1} \: 
	 \:\cos{\left(\frac{\pi}{2}
	(\sqrt{2} \cartu{3} + 1 )\right)}  \\
	&+\; \sigma_1\:\sigma_2\:\cartu{2} \exp{\left(s(\cartusq{1} + 
	\cartusq{2})\right)}
	\:\sin{\left(\frac{\pi}{2}
	(\sqrt{2} \cartu{3} + 1 )\right)} \Big].  
\end{align*}
We implement S3 for a set of objective functions. The chosen set consists
of two-dimensional nodal basis functions along the $\theta$ and $\phi$ spherical coordinate
axes. The goal is to compute the derivative with respect to $s$ of the long-time average 
of each of the nodal basis functions. The Kuznetsov-Plykin map has a one-dimensional 
unstable subspace justifying the use of the algorithm listed in section \ref{sec:algorithm}.
In Fig.1(b), we show the sensitivities computed using the S3 algorithm in section 
\ref{sec:algorithm}, along a trajectory of length 100,000. In comparison, 
a second-order accurate finite difference approximation obtained on 
10 billion samples on the attractor, is used to compute the sensitivities shown in Fig.1(a). 
From the figure, we observe that the S3 sensitivities agree closely with the finite difference sensitivities,
thus validating our approach.

\section{Conclusion}
The problem of sensitivity computation in chaotic systems is under active investigation 
as of this writing. In particular, the problem is the determination of the derivative 
with respect to control or design inputs, of a long-time average or the ensemble mean of 
objective functions of interest in statistically stationary turbulent fluid 
flows. The current methods based on shadowing \cite{angxiu,patrick} and those based on  
ensemble averaging of linearized sensitivities \cite{lea} suffer from lack of 
consistency and prohibitive computational cost, respectively. In this work, we propose 
the space split sensitivity (S3) algorithm as a potential solution to this problem that 
circumvents the pitfalls of these previous approaches. 

In the S3 algorithm, the stable and unstable contributions in Ruelle's \cite{ruelle}
formula, are separated. A stable tangent equation is used to approximate the stable 
contribution as an ergodic average with Monte Carlo convergence. The unstable contribution 
is modified into a time-correlation which can again be computed efficiently through 
a Monte Carlo method. The algorithm requires splitting the tangent vector corresponding 
to the parameter perturbation into its stable and unstable components, at each phase 
point along a trajectory. This direct sum decomposition is achieved from the knowledge 
of the CLV basis for the tangent and adjoint unstable subspaces at each trajectory point. 
In this work, we develop the S3 algorithm under the simplifying 
assumption of a one-dimesional unstable subspace. We validate S3 on the Kuznetsov-Plykin 
attractor, an example of a uniformly hyperbolic system with a one-dimensional unstable 
subspace; comparison against a brute force finite difference approach shows close agreement.

\section*{Appendix}
\subsection{Tangent and adjoint CLV bases}
\label{sec:normalization}
We use $V^1(u),\cdots,V^d(u)$ to denote the tangent CLV basis and 
$W^1(u),\cdots,W^d(u)$ to denote the adjoint CLV basis. The indexing 
of the vectors is such that $V^i,W^i$ correspond to the LE $\lambda^i$ 
and we have the order $\lambda^1 > \lambda^2 \cdots > \lambda^d$. Fixing
an initial condition $u_0$, we denote these bases vectors along the 
trajectory $\left\{ u_n\right\}_{n=0}^\infty$ using a subscript $n$. 
That is, $W^i_n := W^i(u_n)$ and $V^i_n := V^i(u_n)$. The bases are 
normalized in the $l^2$ norm on $\mathbb{R}^d$ -- $\norm{V^i_n} = 
\norm{W^i_n} = 1$, for all $i=1,\cdots,d$ and $n=0,1,\cdots$. 
From the covariance property of the CLVs, the following relationships 
hold -- $F_{un} V_n^i = \norm{F_{un} V_n^i} V_{n+1}^i$ and  
$F_{un}^T W_{n+1}^i = \norm{F_{un}^T W_{n+1}^i} F_{un}^T W_n^i$.

We can decompose an arbitrary tangent vector $Y_n \in T_{u_n} {\cal M}$ into its 
stable and unstable components, with only the unstable tangent and adjoint CLVs. This is possible by making
use of the TAO property (see section \ref{sec:tao}). To see that, suppose $Y_n 
= y_n V^1_n + Y_n^{\rm s}$, where $Y_n^{\rm s} \in E^{\rm s}(u_n)$. The TAO property 
says that $Y_n^{\rm s} \cdot W^1_n = 0$. Therefore, $y_n = Y_n \cdot W^1_n /c_n^1$, 
where $c_n^1 := V_n^1\cdot W_n^1$. 

\subsection{Computable formula for the unstable contribution}
\label{sec:unstableContributionFormula}
We are interested in computing the following summation
\begin{align}
	\sum_{n'=0}^\infty 
	\langle \grad{J_{n'}}{0}\cdot X_0^{\rm u}, \mu^s \rangle,  
\end{align}
which, for large $n'$, leads to an inefficient computation through ergodic averaging. Denoting
by $x_n$ the component of $X^{\rm u}_n$ along the unstable CLV at $u_n$, we can write the 
term we wish to compute as,
\begin{align*}
	\sum_{n'=0}^\infty 
	\langle x_0 \grad{J_{n'}}{0}\cdot V_0^1, \mu^s \rangle =  
	\sum_{n'=0}^\infty \Big(
	\langle \grad{x_0J_{n'}}{0}\cdot V_0^1, \mu^s \rangle - 
	\langle  J_{n'} \grad{x_0}{0}\cdot V_0^1, \mu^s \rangle \Big).
\end{align*}
This amounts to doing integration by parts. The second term can be computed
efficiently as an ergodic average. Using the measure-preserving property of $F$, 
we can write the first term as,
\begin{align*}
	\sum_{n'=0}^\infty 
	\langle \grad{x_{0}J_{n'}}{0}\cdot V_{0}^1, \mu^s \rangle = 
	\sum_{n'=0}^\infty 
	\langle \grad{x_{1}J_{n'+1}}{1}\cdot V_{1}^1, \mu^s \rangle = 
	\sum_{n'=0}^\infty 
	\langle F_{u0}^{-T} \grad{x_{1} J_{n'+1}}{0}\cdot V_{1}^1, \mu^s \rangle,
\end{align*}
where we have used chain rule to rewrite the derivative term. Thus we have, 
\begin{align*}
	\sum_{n'=0}^\infty 
	\langle \grad{x_0 J_{n'}}{0}\cdot V_0^1, \mu^s \rangle &= 
	\sum_{n'=0}^\infty 
	\langle \grad{x_{1}J_{n'+1}}{0}\cdot F_{u0}^{-1} V_{1}^1, \mu^s \rangle 
	= \sum_{n'=0}^\infty 
	\langle \grad{x_{1}J_{n'+1}}{0}\cdot \frac{V_{0}^1}{\norm{F_{u0} V_{0}^1}}, \mu^s \rangle, 
\end{align*}
where to obtain the last equality, we have used the normalization relationships of 
CLVs from section \ref{sec:normalization}. We can therefore express the quantity of 
interest as follows by rewriting the derivative,
\begin{align}
	\label{eqn:iterationBase}
	\sum_{n'=0}^\infty 
	\langle \grad{x_0J_{n'}}{0}\cdot V_0^1, \mu^s \rangle &= 
	\sum_{n'=0}^\infty \Big( 
	\langle \grad{\frac{x_{1}J_{n'+1}}{\norm{F_{u0}V_{0}^1}}}{0}\cdot V_{0}^1, \mu^s \rangle 
	- \langle x_{1} J_{n'+1} \grad{\frac{1}{\norm{F_{u0} V_{0}^1}}}{0}\cdot V_{0}^1, \mu^s \rangle 
	\Big).
\end{align}
The above equation is also true if we replaced the objective function with a 
scaled form of itself. Let the scaled objective function be 
$\tilde{J}_0 = J_1 x_{-n'+1}/(x_{-n'}\norm{F_{u(-n')}V_{-n'}^1})$. 
Substituting $\tilde{J}$ in place of $J$ in Eq. \ref{eqn:iterationBase},
\begin{align}
	\label{eqn:iterationPullback}
	\sum_{n'=0}^\infty 
	\langle \grad{\frac{x_{1}J_{n'+1}}{\norm{F_{u0}V_{0}^1}}}{0}\cdot V_0^1, \mu^s \rangle &= 
	\sum_{n'=0}^\infty \Big( 
	\langle \grad{\frac{x_2J_{n'+2}}{ \norm{F_{u1}V_1^1} \norm{F_{u0}V_0^1}}}{0}\cdot V_0^1, \mu^s \rangle 
	- \langle \frac{x_2 J_{n'+2}}{\norm{F_{u1}V_1^1}} \grad{\frac{1}{\norm{F_{u0}V_0^1}}}{0}\cdot V_0^1, \mu^s \rangle 
	\Big). 
\end{align}
Note that the LHS of Eq.\ref{eqn:iterationPullback} is the same as the first term of the 
RHS of Eq.\ref{eqn:iterationBase}. Thus, we can rewrite Eq.\ref{eqn:iterationBase} as,
\begin{align}
	\label{eqn:1iteration}
	\sum_{n'=0}^\infty 
	\langle \grad{x_0J_{n'}}{0}\cdot V_0^1, \mu^s \rangle &= 
	\sum_{n'=0}^\infty \Big(
	\langle \grad{\frac{x_2J_{n'+2}}{ \norm{F_{u1}V_1^1} \norm{F_{u0}V_0^1}}}{0}\cdot V_0^1, \mu^s \rangle 
	- \langle \frac{x_2 J_{n'+2}}{\norm{F_{u1}V_1^1}} \grad{\frac{1}{\norm{F_{u0}V_0^1}}}{0}\cdot V_0^1, \mu^s \rangle \\ 
	&- \langle x_{1} J_{n'+1} \grad{\frac{1}{\norm{F_{u0} V_{0}^1}}}{0}\cdot V_{0}^1, \mu^s \rangle 
	\Big).
\end{align}
By replacing the objective function in Eq.\ref{eqn:iterationBase} with 
$\tilde{J}_0 = J_2 x_{-n'+2}/(x_{-n'}\norm{F_{u(-n')}V_{-n'}^1} \norm{F_{u(-n'+1)}V_{-n'+1}^1})$, 
we can continue the recursion. With every step of the recursion, the objective function decreases 
by a factor > 1 and thus goes to 0, as the number of recursive steps tends to $\infty$. Hence, our 
desired term reduces to the following summation over $n$ (the number of such recursive steps),
\begin{align}
	\label{eqn:finalIteration}
	\sum_{n'=0}^\infty 
	\langle \grad{x_0J_{n'}}{0}\cdot V_0^1, \mu^s \rangle &= 
	-\sum_{n'=0}^\infty \sum_{n=1}^\infty 
	\langle \frac{x_n J_{n'+n}}{\Pi_{i=1}^{n-1}\norm{F_{ui}V_i^1}} 
	\grad{\frac{1}{\norm{F_{u0}V_0^1}}}{0}\cdot V_0^1, \mu^s \rangle.
\end{align}
For the sake of brevity, let us define $\phi_n := -\grad{1/\norm{F_{un} V_n^1}}{n}\cdot V_n^1$. Applying 
measure preservation and using the more compact notation, we can rewrite Eq.\ref{eqn:finalIteration} as 
\begin{align}
	\label{eqn:finalIterationCompact}
	\sum_{n'=0}^\infty 
	\langle \grad{x_0J_{n'}}{0}\cdot V_0^1, \mu^s \rangle &= 
	\sum_{n'=0}^\infty \sum_{n=1}^\infty 
	\langle \frac{x_0 J_{n'}\phi_{-n}}{\Pi_{i=1-n}^{-1}\norm{F_{ui}V_i^1}}, \mu^s \rangle.
\end{align}

The integrand is bounded at all $n$ and the summation over $n$ is absolutely convergent. Thus the summation over 
$n$ and the ensemble average can be commuted to yield,
\begin{align}
	\label{eqn:reduceSummation}
	\sum_{n'=0}^\infty 
	\langle \grad{x_0J_{n'}}{0}\cdot V_0^1, \mu^s \rangle &= 
	\sum_{n'=0}^\infty
	\langle x_0 J_{n'} \sum_{n=1}^\infty \left( \frac{\phi_{-n}}{\Pi_{i=1-n}^{-1}\norm{F_{ui}V_i^1}} \right), 
	\mu^s \rangle.
\end{align} Thus, we have essentially circumvented the computation of the derivative of $J_n'$. This was made 
possible by writing,
\begin{align}
	\label{eqn:reduceSummation}
	\sum_{n'=0}^\infty 
	\langle \grad{x_0J_{n'}}{0}\cdot V_0^1, \mu^s \rangle &= 
	\sum_{n'=0}^\infty
	\langle J_{n'} \alpha^{\rm u}_0, \mu^s \rangle, 
\end{align}
where $\alpha^{\rm u}_0 :=  x_0 \sum_{n=-\infty}^{-1} \frac{\phi_n}{\Pi_{i=1+n}^{-1}\norm{F_{ui}V_i^1}}$ is a scalar field
that can be computed efficiently.
\section*{Acknowledgments}
This work was supported by AFOSR Award
FA9550-15-1-0072 under Dr. Fariba Fahroo and Dr. Jean-luc Cambrier.
The authors thank Benjamin Zhang for comments on this manuscript.
\bibliography{sample}

\end{document}